\documentstyle[aps,prl]{revtex}

\begin{document}
\twocolumn[
{\hsize\textwidth\columnwidth\hsize\csname@twocolumnfalse\endcsname
\draft
\title{Experimental evidence for a spin gap in the s=1/2 quantum antiferromagnet Cu$_2$(OH)$_2$CO$_3$}
\author{E. Janod$^{(1)}$, L. Leonyuk$^{(2)}$ and V. Maltsev$^{(1,2)}$}
\address{$^{(1)}$ Institut des Materiaux Jean Rouxel,Universite de Nantes-CNRS, 44322 Nantes Cedex 3, France}
\address{$^{(2)}$ Moscow State University, Moscow 119899, Russia}
\date{\today}
\maketitle

\begin{abstract}
Magnetic properties of the natural mineral Cu$_2$(OH)$_2$CO$_3$ (malachite) were investigated through DC and AC susceptibility measurements. 
The analysis of the low-temperature part reveals a quantum spin-gap behavior with $\Delta \approx 130 K$. Consistently with the crystal structure, the magnetic susceptibility can be accurately described by a model of alternating chain. The non-frustrating residual inter-chain magnetic couplings, describing a sort of dimerized square planar lattice, are not strong enough to push the system towards a long-range ordered ground state, in good agreement with recent theoretical studies.
\end{abstract}

}
]

An intense renewal of interest for low dimensional quantum spin systems has stemmed from the Haldane's conjecture \cite{1}, which pointed out the essential difference between the ground state properties of the antiferromagnetic Heisenberg (AFH) chain with integer and half-integer spin. Quantum fluctuation may prevent low-dimensional AFH systems from falling in a long-range ordered (LRO) Néel-like ground state for particular topologies of magnetic interactions and spin values. For instance, Haldane proposed that the uniform spin 1 chain would exhibit a quantum disordered (spin liquid) ground state (Haldane phase) with a gap in the spin excitation spectrum, whereas the uniform spin 1/2 chain would be gapless and quasi-LRO. There has been considerable evidence in support of this statement during the last two decades on both experimental \cite{2} and theoretical sides \cite{3}. Interestingly, the uniform AFH spin ½ chain is in a critical state, in the sense that an infinitesimal perturbation may push the system into a completely different ground state. In the case of exchange dimerization, the resulting system is an alternating chain which has a gapped disordered ground state with a string order closely related with the Haldane phase's one \cite{4}.

One of the current issues in the field of quantum magnetism is the occurrence of the spin liquid state in two-dimensional (2D) lattices, with or without the presence of magnetic frustration. In non-frustrated 2D AFH systems, the case of dimerized square planar lattices, built up from alternating chain connected by inter-chain coupling, is especially interesting. With a vanishing dimerization within the chain and a strong enough inter-chain coupling, the uniform square planar, which ground state is LRO, can be recovered. On the other hand, a zero inter-chain coupling and a finite dimerization leads to a simple alternating chain, with a disordered spin liquid ground state. A quantum critical line therefore separates the two different ground states in the dimerization – inter-chain coupling sub-space. Several theoretical studies have been devoted to this subject \cite{5,6,7,8}.

In this context, we studied the magnetic properties of the malachite Cu$_2$(OH)$_2$CO$_3$. This well-known mineral, also used in jewelry, has, to our knowledge, never been investigated for its quantum magnetic properties. The crystal structure is monoclinic (space group P12$_1$/$a$1) with cell parameters $a$=9.502 \AA, $b$=11.974 \AA, $c$=3.240 \AA and $\theta$=98.75$^{o}$
 \cite{9}. The structure is built up from corrugated copper oxides ($ac$) planes formed with edge and corner-sharing distorted CuO$_6$ octahedra (see Fig. 1). The CO$_3$ groups bridge the neighboring planes together and ensure the three-dimensional stability. The copper electronic configuration is $3d^9$ ($Cu^{2+}$, s=1/2). In this paper, we show that the dimensionality of the magnetic interaction lattice defined by the different magnetic couplings between the spins 1/2 is intermediate between 1D and 2D in Cu$_2$(OH)$_2$CO$_3$. Taking into account the four principal magnetic interactions, the structure appears to be quasi-2D consisting, within the $ac$ plane, in a set of parallel alternating chains connected by weak inter-chain coupling, resulting in a kind of dimerized square planar lattice. Despite the inter-chain couplings, the magnetic susceptibility can be accurately described using the high precision fitting procedure recently developed by Johnston and co-workers \cite{10} for an quantum disordered alternating chain. This clearly indicates that inter-chain couplings are too weak to push the system towards a LRO ground state, as expected from recent theoretical studies \cite{5,6,7,8}.

\section{ Experimental}

Two different natural samples of Cu$_2$(OH)$_2$CO$_3$, originating one from the Urals, Russia and one from Zaire, were used in this study. Their purity was checked by X-rays diffraction and Electron Probe Microscope Analysis. The samples were found to be single phased and no impurities were detected.
DC and AC magnetic measurements were performed in a commercial Quantum Design MPMS-5S SQUID magnetometer. The fields were set to 1 kOe for the DC measurement and to 2 Oe for the AC one, with a driving frequency of 997 Hz.

The rather complex fitting formula for the magnetic susceptibility of an alternating chain proposed by Johnston and co-workers \cite{10} was inserted in a program including a Simplex fitting algorithm. We checked the absence of any error among the eighty four constants required in the formula by fitting successfully numerical data obtained for alternating chains by the transfer-matrix density-matrix renormalization group (TMRG) method \cite{10} for $\alpha$ =0 (dimer), 0.5, 0.95 and 1 (uniform spin 1/2 chain).

\section{ Results}
Figure 2 shows the magnetic susceptibility data measured at 1 kOe. The most striking feature is the presence of a broad maximum in the $\chi(T)$ curve at 120 K. This behavior is typical either of low dimensional antiferromagnets or of systems with a gap in the spin excitation spectrum. The drop of the susceptibility below the broad maximum is balanced below 25 K by a Curie-like tail. In order to detect a possible spin gap, we performed a low temperature fit which describes the susceptibility as the sum of a T-independent $\chi_o$ term plus a Curie-Weiss $C/(T-\theta)$ law and a thermally activated susceptibility $A T^{-1/2} e^{-\Delta/T}$ suitable for most of the quasi-1D spin gap systems \cite{11}. The good quality of the fit, depicted in the inset of Fig. 2, strongly supports the existence of a spin gap in the malachite. The low temperature tail corresponds to 2.4 percent of free spin ½ per Cu. The value of the spin gap loosely depends on the fitted temperature range (typically 2-60 K) and stay within the 110-130 K range. Interestingly, it roughly corresponds to the position of the maximum of the susceptibility $T(\chi_{max})$, implying a ratio $\Delta/T(\chi_{max})$ close to unity. As a consequence, the malachite is probably not a trivial collection of isolated dimers of spin 1/2, for which the ratio $\Delta/T(\chi_{max})$ is around 1.6. We also attempted to fit our data using the expression for isolated dimers, with the number of dimers as a free parameter. For any reasonable value of the g factor (2 $<$ g $<$ 2.25), it leads to an unrealistic value of number of dimers, between 60 and 70 percent of the expected value.

As we will see below, magnetic frustration could possibly exist in the malachite. We therefore measured the AC susceptibility $\chi_{AC}$=$\chi'$+i$\chi''$ of the malachite to detect a possible spin glass-like behavior at low temperature. It results (i) in a perfect matching between the component $\chi'$ and the DC susceptibility and (ii) in the absence of any peak in the $\chi''$ part down to 2 K, consistent with the absence of any ordered state down to low temperature.

\section{ Discussion}
In order to get more insight into the real magnetic interaction network of the malachite, it is important to establish at least a hierarchy of the different magnetic couplings existing in this system. The leading magnetic coupling here originates in Cu-O-Cu superexchange. The main parameters affecting the strength and the sign of superexchange are the Cu-O distance, the Cu-O-Cu angle and the crystal structure via the Madelung potential \cite{12}. The Anderson-Kanamori-Goodenough rules \cite{13} state that the super-exchange Cu-O-Cu is the sum of a weak angle-independent ferromagnetic coupling and of a strong antiferromagnetic coupling maximum for $\phi$= 180$^{o}$ and tending to zero at 90$^{o}$. The total magnetic interaction is therefore expected to cross zero and to change its sign close to 90$^{o}$. A theoretical approach based on a three bands Hubbard model and dedicated to the study of several cuprates with edge-sharing CuO$_{4}$ units showed that this crossing occurs close to $\phi$ = 95$^{o}$ \cite{14}. 

The above description of the structure in terms of corrugated planes of octahedra is somewhat misleading since most of these octahedra have a broad distribution of copper-oxygen distances ranging from 1.89 \AA to 2.67 \AA   \cite{9}. By restricting the pertinent Cu-O distances to the 1.89-2.11 \AA range, all the polyhedra change from CuO$_6$ octahedron to distorted CuO$_4$ squares planar and the 2D connectivity is lost within the ac plane, as shown in Fig. 3(a). The effective magnetic network only consists in parallel chains running in the ac plane and by only two types of magnetic interactions J$_1$ and J$_2$ alternating along the chains (see Fig. 3(b)). A detailed description of J$_1$ and J$_2$ is given in Table I. By comparison with the phase BaCu$_2$Si$_2$O$_7$ \cite{16}, where almost similar Cu-O-Cu bonds lead to J=280 K, a rough estimate gives 200 $\leq$ J$_1$ $\leq$ 400 K. As the coupling J$_2$ is mainly mediated through a Cu-O-Cu bond with an angle of 106$^{o}$ (122$^{o}$ for J$_1$), J$_2$ must be significantly smaller than J$_1$.

By decreasing order of intensity, the next magnetic couplings should be located within the ac planes and link the chains together. Three new magnetic inter-chain couplings J$_3$, J$_4$ and J$_5$ (see Fig. 3(c) and Table I) can be distinguished and are all mediated via Cu-O distances close to 2.37 \AA. J$_5$ should be negligible compared to J$_3$ and J$_4$, since the involved Cu-O-Cu angles, 97.5$^{o}$ and 97.7$^{o}$, are very close to the critical value (95$^{o}$ in the cuprates) for which the magnetic superexchange vanishes \cite{14}. The network including the four main interactions J$_1$, J$_2$, J$_3$ and J$_4$, shown in Fig. 3(c), has strong similarities with the non-frustrated dimerized square lattice treated theoretically by Singh and co-workers [5] and depicted in Fig. 3(e). Estimating the J$_3$ and J$_4$ values is not a simple task; the Cu-O-Cu angles in J$_3$ and J$_4$ are however close to the one appearing in J$_1$ and J$_2$, respectively (see Table I). Therefore the main difference originates in the Cu-O distances (2.37 \AA against 1.91 \AA). Assuming a power-law dependence $r^{-n}$ of the superexchange strength with distance, a value n=10-11 can be inferred from the work of Mizuno et al. \cite{14} on Cu-O networks. As $(2.37/1.91)^{-11}$ $\sim$ 0.1, J$_3$ and J$_4$ can be estimated to be roughly one order of magnitude smaller that J$_1$ and J$_2$, respectively. Note that, contrary to J$_3$ and J$_4$, the interaction J$_5$ could lead to a magnetic frustration since it couples spins over the diagonal of the dimerized square.

The last kind of magnetic coupling that can be considered are those mediated by the CO$_3$ groups. It results in (i) a next nearest neighbor (NNN) interaction within the J$_1$-J$_2$ chain and (ii) in a set of inter-plane couplings J$_{IP}$, inducing a strong frustration as shown in Fig. 3(d).
Finally, Dzyaloshinsky-Moriya interactions could possibly exist in the malachite because of the absence of inversion center between the different neighboring Cu \cite{17}. They are however most probably negligible since the magnitude of the low temperature susceptibility is hardly reconcilable with the existence of a weak ferromagnetism.

From the above magneto-structural analysis, it appears that a simple model of alternating chains is a good starting point to understand the magnetism of the malachite. As previously discussed, the ratio $\Delta/T(\chi_{max})$ is close to one in the malachite. In an alternating chain where strong bonds $J$ alternate with weaker bonds $\alpha J$ ($\alpha$<1), such a feature is expected for $\alpha$ close to 0.5 \cite{10}. Also, as $T(\chi_{max})/J$ loosely depends on the dimerization $\alpha$ in an alternating chain ($T(\chi_{max})/J$ $\sim$ 0.6 \cite{10}), the deduced value of the strongest magnetic coupling within the chain in the malachite is J$_1$ $\sim$ 200 K. This corresponds to the lower limit of  J$_1$ estimated from magneto-structural arguments.
To be more quantitative, we tried to fit the susceptibility curve of our two samples using a fitting procedure for alternating chains recently proposed in ref. [10] and suitable for any dimerization parameter 0 $\leq$ $\alpha$ $\leq$ 1. We used the formula:

\begin{eqnarray}
 \chi(T)&=&\chi_{o}+C/(T-\theta) \nonumber \\
&&+Ng^{2}\mu_{B}^{2}/(k_{B}J)*\chi_{alt. chain}(T,J,\alpha)
\end{eqnarray}

where $\chi_o$ is a T-independent term, C the Curie constant, $\theta$ the Weiss temperature and $J$ the strongest coupling within the alternating chain. The free parameter were $\chi_o$, $C$, $\theta$, $g$, $J$ and $\alpha$. The best result of the fit, depicted in Fig. 2, was obtained for $g$ = 2.08 (2.20), $J$ = 196K (203 K) and $\alpha$ = 0.49 (0.53). The values in brackets correspond to the results obtained on the second sample (not shown in Fig. 2). The most important parameters of the fit, $J$ and $\alpha$, appear to be very similar from one sample to another. The resulting spin gap can be computed using the formula $\Delta$/$J$ $\approx$ (1+$\alpha$)$^{1/4}$ (1-$\alpha$)$^{3/4}$  \cite{10} and is equal to 131 (128) K. This value compares remarkably well with the one obtained from the low-temperature approach ($\Delta$=126 K).

The good agreement between data and the global fit proposed by Johnston and co-workers supports the model of simple alternating chains. However, a fit of the magnetic susceptibility is not sufficient to prove that the magnetic network considered is correct. It is therefore important to check if a quantum spin-gap state is expected taking into account the presence of (i) the possible NNN coupling within the chain, (ii) the inter-chain coupling J$_3$ and J$_4$. 

The coupling induced by the CO$_3$ groups can safely assumed to be small. It would otherwise result in a 3D highly frustrated magnetic lattice, which is hardly reconcilable with the existence of a large spin gap in the malachite. As far as the inter-chain couplings J$_3$ and J$_4$ are concerned, it is possible to be more quantitative. Singh and co-workers \cite{5} considered the topology depicted in Fig. 3(e), similar as the dimerized square planar description of the malachite but with J$_2$=J$_3$=J$_4$=J'. They found a quantum critical point for $\alpha_{c}$=(J'/J)$_c$=0.39 $\pm$ 0.01, separating the LRO ground state ($\alpha$ $>$ $\alpha_c$) from the spin liquid disordered one ($\alpha$ $<$ $\alpha_c$). Qualitatively, it is expected that $\alpha_c$ will tend to the critical value of a pure alternating chain ($\alpha_c$=1) in the limit of small inter-chain (J$_3$,J$_4$) coupling, which is the case in the malachite. This intuition is confirmed by the work of Katoh and Imada \cite{7}. They considered a slightly different dimerized square planar topology shown in Fig. 3(f), where strong bonds J alternate with weak bonds $\alpha J$. As the inter-chain coupling J$_3$=J$_4$=J$_{\perp}$ decreases from J$_{\perp}$/J=0.70 to 0.26, the critical dimerization $\alpha_c$ increases from 0.40 to 0.74. 

In the malachite, our previous estimate of the ratio inter-chain over intra-chain couplings, J$_{\perp}$/J $\approx$ J$_3$/J$_1$, and of the dimerization parameter $\alpha$ were J$_{\perp}$/J $\approx$ 0.1 and $\alpha$ $\approx$ 0.50, respectively. From the above discussion, it can be inferred that the critical dimerization $\alpha_c$ for the malachite is larger than $\alpha$ $\approx$ 0.50 for an inter-chain coupling J$_{\perp}$/J $\approx$ 0.1. A disordered spin liquid ground state is therefore expected in the malachite, in agreement with the low temperature fit of the susceptibility.

\section{Conclusion}
In summary, the natural mineral Cu$_2$(OH)$_2$CO$_3$ exhibits a quantum spin liquid ground state, consistent with its structure including s=1/2 alternating chain with rather low inter-chain couplings. Also, the malachite provides a good example of interesting magnetic properties in a natural mineral. In the field of quantum magnetism, other copper-oxide based minerals could possibly be a rich source of new materials that has only been poorly investigated so far. 
\\
\\
Acknowledgments
\\We would like to thank C. Payen for stimulating discussions and for the critical reading of the manuscript. We also thank the Fersman Mineralogical Museum of the Russian Academy of Science for providing us one of the natural malachite sample.

\begin{figure}[tbp]

\caption{Crystal structure of the malachite Cu$_2$(OH)$_2$CO$_3$. The main structural units are the CuO$_6$ octahedra and the CO$_3$ triangles.}

\label{fig1}

\end{figure}

\begin{figure}[tbp]

\caption{DC magnetic susceptibility of Cu$_2$(OH)$_2$CO$_3$. Data are depicted in open squares, the fit of the susceptibility following the model of alternating chains as a solid line and the intrinsic susceptibility $\chi_{measured}-\chi_o-C/(T-\theta)$ as a dashed line. The inset shows the same data in a double-logarithm plot as well as the low temperature fit, as described in the text.}

\label{fig2}

\end{figure}

\begin{figure}[tbp]

\caption{(a) Copper-oxygen lattice obtained by keeping only the Cu-O distances below 2.11 \AA.(b) Equivalent magnetic lattice,  where magnetic couplings are represented as solid and dashed segments.(c) Magnetic lattice including the four leading couplings. (d) Details of the interactions within (along the a axis) and perpendicular to (along the b axis) the chains, including those possibly mediated by the CO$_3$ groups (see text). (e) and (f) Magnetic topologies considered in Ref. [5] and Ref. [7], respectively.}

\label{fig3}

\end{figure}
\end{document}